\documentclass[pra,aps,twocolumn,showpacs]{revtex4}
\usepackage{graphicx}
\begin{document}
\title{Optical phase shifts 
and diabolic topology in M\"{o}bius-type strips} 
\author{ Indubala I Satija$^1$ and Radha Balakrishnan$^2$  }
\affiliation{(1) Department of Physics,George Mason University, Fairfax, VA
22030.\\
(2)The Institute of Mathematical Sciences, Chennai  600 113,
India\\}
\begin{abstract}

We compute the optical phase shifts between the left and the right-circularly polarized
light after it traverses non-planar cyclic paths described by the boundary curves of 
closed twisted strips.
The evolution of the electric field along the curved path of a light ray is 
described by the Fermi-Walker 
transport law  which is mapped to  a Schr\"{o}dinger equation.
The effective quantum Hamiltonian of the system has
eigenvalues equal to  $0, \pm \kappa$, where $\kappa$ is the local
curvature of the path. The inflexion points of the twisted strips
correspond to the vanishing of the curvature and 
manifest themselves as the diabolic crossings of the quantum Hamiltonian.
For the M\"{o}bius loops,
the critical width where the diabolic geometry resides also corresponds to the 
characteristic width
where the optical phase shift is  minimal.
In our detailed study of various
twisted strips, this intriguing property singles out the M"{o}bius  geometry.

\end{abstract}
\maketitle
The geometrical phenomenon of anholonomy relates to the inability of a variable to return to
its original value after a cyclic evolution.\cite{book} An example of this phenomenon is the 
change in the direction of polarization of light in a coiled optical 
fiber.\cite{Chiao, tomita} This leads to an 
optical phase shift which
corresponds to a phase change between a right and a left-handed
circularly polarized wave, when they travel along a non-planar path.
\cite{optics} 
The  effect is an optical manifestation of the Aharonov-Bohm phase, according to which two
electron beams develop a phase shift proportional to the magnetic flux they enclose.\cite{AB}
For the polarized light the analog of magnetic flux is the solid angle subtended on the
sphere of directions $\hat{{\bf k}}$, where $\hat{{\bf k}}$ represents the direction of 
propagation of light, which changes
as light passes through a twisted fiber.
In previous studies  involving the propagation of light through a helically wound optical fiber\cite{Chiao, tomita, optics},
the simple geometry of a circular helix gives this solid angle to be equal to 
$\Omega=2\pi (1-\cos~ \theta)$,
where $\theta$ is the pitch angle of the helix, or the angle between the axis of the helix and 
the local axis of the optical fiber.

In this paper, we compute the optical phase shift between the left and the right-circularly polarized
light as it passes through the optical fibers with a more complex geometry,
namely, fibers shaped like the {\it boundary curves} of closed  twisted strips. Although we study a whole class of twisted strips,
our main focus is on the M\"{o}bius strip whose unique geometry has been a source of constant fascination.
Here we explore the interplay between the intrinsic geometry of the strips and the 
geometrical
physical phenomenon, namely the geometric phase shift experienced by light as it passes
through the boundary curve of the strips. Interestingly, the M\"{o}bius geometry is singled
out, since the optical phase shift is found to exhibit a characteristic minimum as the 
width of the strip is continuously varied.
 
The class of twisted, closed strips under consideration is described by,
\begin{eqnarray}
x&=&(1+w \cos\frac{nt}{2})\cos t;\nonumber\\
y&=&(1+w \cos\frac{nt}{2})\sin t;\nonumber\\
z&=&w\sin\frac{nt}{2},\nonumber\\
\label{ms}
\end{eqnarray}
with parameters $-\alpha\le w \le \alpha $ and $0\le t \le 2\pi$.
Thus $\alpha$ is the half-width of the strip and its length $L=2\pi$.
The integer $n$ is the number of half-twists on the strip (so that
$n=1$ refers to the well-known M\"obius strip). 
We are interested in the geometry and topology of  
  the boundary curve of the twisted strip. The  parametric
equation for this space curve is found by setting  
 $w=\pm \alpha$ in Eq. (\ref{ms}).
For odd $n$, the strip is a  non-orientable surface with one boundary
curve, so that the range of $t$ is $[0,4\pi]$. 
For even $n$, the strip is orientable, 
with two boundary curves
with similar geometries, and 
the range of $t$ is $[0,2\pi]$ for each of these curves.
 Topologically, for all widths, the 
boundaries of  the odd-$n$ strips  with $n > 1$
are knotted curves 
(e. g., it is a trefoil knot for $n=3$, a 
five-pointed star knot for $n=5$, etc.), 
while those of the even-$n$ 
strips are not knotted. The boundary curve of the 
M\"{o}bius strip clearly does not fall 
in either of these classes, since it is the only
case in which the  boundary curve
 of a non-orientable strip is not a knot. 
Figure ~1 shows the boundary curves of the M\"{o}bius strip
for various values of  $\alpha$. 
One of the characteristic feature of the twisted strips are the
inflexion points, distinguished by a local "straightening" around that point.
Various details regarding such points for $n$=twisted strips and
their non-trivial dependence
on the width-to-length ratio will be discussed later. An important point of this paper
is that for the M\"{o}bius strip, the inflexion point manifests itself in a rather unique way
in encoding the polarization properties of light as it traverses the M\"{o}bius loop.

We consider the propagation of circularly polarized light along an optical fiber which has the 
shape of the boundary curve 
of a twisted strip.
The direction of propagation of light is tangential to the boundary curve and hence completely encodes the
geometry of the boundary curve. As the light completes a full loop of the twisted strip
boundary, its direction of propagation, the tangent indicatrix, completes a closed circuit 
on the surface of the sphere of directions defined by the propagation 
vector $\hat {\bf k}$.
By adiabatically varying the direction of propagation around a closed circuit on the sphere,
the change in the polarization of light is equal to the solid angle subtended in ${\bf k}$-space.
Condition of adiabaticity is that the length of the boundary curve be large as compared to the 
wavelength of light.

The boundary curve of a twisted strip,  viewed
as a space curve $\Gamma:{\bf r}(s)=(x,y,z)$, thus 
represents the path along which light propagates.
Here $s$ is the arc length measured along the curve, with $ds=v~dt$, where 
$v=|\frac{d\bf{r}}{dt}|$.
The geometry of the space curve can be described by the right-handed orthonormal triad
(${\bf T}, {\bf N}, {\bf B}$) that represent unit tangent, normal and binormal
at every point along the trajectory. The evolution of the triad on the curve can
be described by Frenet-Serret equations\cite{FS},
\begin{equation}
\frac{d{\bf T}}{ds}=\kappa {\bf N};~
\frac{d{\bf N}}{ds}=-\kappa {\bf T}+\tau {\bf B};~
\frac{d{\bf B}}{ds}=-\tau {\bf N},
\label{FS}
\end{equation}
where ${\bf B}={\bf T} \times {\bf N}$. 
$\kappa$ and $\tau$ denote, respectively, the local curvature and the 
torsion. They are given by\cite{FS} 
\begin{equation}
\kappa =|{\bf T}(s)|~~~;~~~ 
\tau~=~[{\bf T}\cdot \frac{d{\bf T}} {ds} \times \frac{d^2{\bf T}}{ ds^2}]/ \kappa^2,
\label{kt}
\end{equation} 
Intuitively, the curvature measures the deviation of the 
curve from a straight line, while the torsion
quantifies the non-planarity of the curve.

In order to compute the change in the polarization of light as it passes 
through a circuit in the shape of 
the boundary of a twisted ribbon, we follow
the variation in the electric field vector with respect to $t$
(or $s$ ) as light travels along the curve $\Gamma$. 
Using Maxwell's equations, it can be shown 
that the  evolution of the complex unit vector 
${\bf \hat{E}}$ associated with the complex three-component electric field
on this space curve follows the Fermi-Walker transport law\cite{BW, MSimon},
\begin{equation}
\frac{d{\bf \hat{E}}}{ds}= \kappa(s) {\bf B} \times {\bf \hat{E}}
\label{SE}
\end{equation}

\begin{figure}[htbp]
\includegraphics[width=2.8in, angle=270]{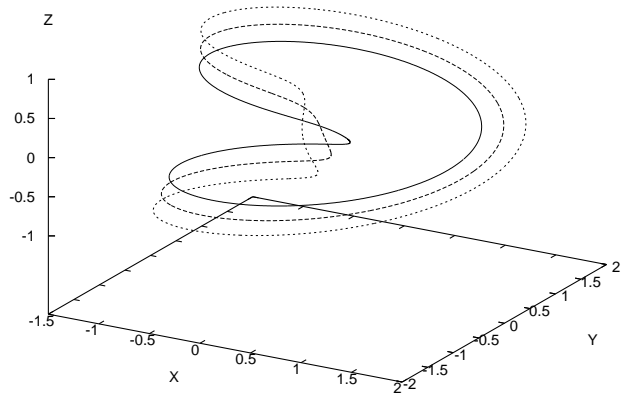}
\leavevmode
\caption{Boundary curves of the closed M\"{o}bius strip 
(Eq. (\ref{ms}) with $n=1$) for strip-widths 
$\alpha=1.0, 0.8, 0.6$ (outermost to innermost). Note  the local straightening
around the inflexion point for the critical curve with $\alpha = 0.8$.}
\label{fig1}
\end{figure}

We first note that interestingly, Eq. (\ref{SE}) can be 
cast in the form of a Schr\"{o}dinger 
equation with the following Hamiltonian
$ H(s) $, which is a three-dimensional, antisymmetric Hermitian matrix, 
with pure imaginary elements.

\begin{equation}
H=i \kappa \left( \begin{array}{ccc} 0 &-B_3& B_2 \\ B_3 & 0 & -B_1\\-B_2 & B_1 & 0\\
\label{H}
\end{array}\right)\\
\end{equation}
Here $B_i$ is the $i^{th}$-component of the binormal vector ${\bf B}$.
A short calculation shows that the eigenvalues of this Hamiltonian are $0, \pm \kappa$.
 Hence the quantum Hamiltonian
exhibits a $3$-fold degeneracy at points where $\kappa$ vanishes. These
are just the inflexion points  of the optical fiber, and this is a general
result that follows from the basic evolution Eq. (\ref{SE}). 
For the example of a fiber shaped like  
the boundary curve of an $n$-twisted strip, there are 
$n$ such inflexion points ( see below) that appear when the width of the strip 
takes on a  certain critical value, which depends on $n$.

\begin{figure}[htbp]
\includegraphics[width=2.7in, angle=270]{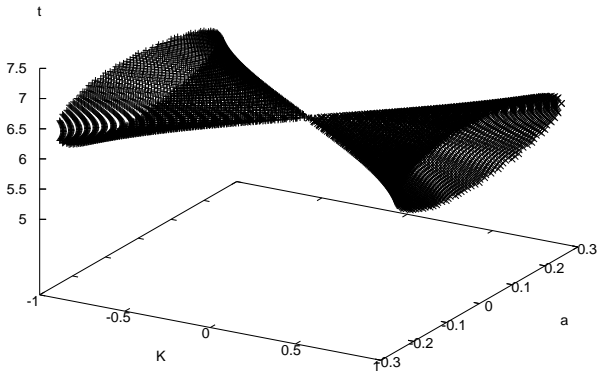}
\leavevmode
\caption{Conical topology of the eigenvalues $\pm \kappa(t,\alpha)$ 
of the quantum Hamiltonian (Eq. (\ref{H}))
near the critical point ($t_c, \alpha_c$) on the surface of 
the M\"{o}bius strip. Here $a=(\alpha-\alpha_c)/ \alpha_c$}.
\label{fig2}
\end{figure}

The vanishing of $\kappa$ implies that the quantum Hamiltonian is 
degenerate at the the critical point and as $t$ and $\alpha$ vary on the surface of the strip,
this degeneracy leads to a diabolic point. Fig. (\ref{fig2})
illustrates the conical geometry
near the critical point. Further details about such points will be discussed later in this paper.
Such diabolic crossings in $t-\alpha$ 
space are sensed by a circuit
that does not pass through the degeneracy, but simply encloses it. In particular, such closed loops
result in wave functions that acquire a  Berry phase
 that depends on the geometry of the path
in {\it parameter space} in  an adiabatic cyclic evolution.
In contrast to Berry phase, which is obtained by 
a mathematical mapping of an optical system to a quantum 
problem, and further, is associated with parametric circuits enclosing
a conical intersection, the geometric phase  
that we investigate involves circuits that are in {\it configuration space},
i.e., boundary curves of twisted strips.  
Hence, while a special circuit may pass through a  diabolic crossing, 
typical circuits never enclose one. 
More importantly, 
this geometric phase can be directly measured 
in experiments, making it more interesting and applicable,
as we shall discuss at the end. 

The geometric phase is obtained by solving  Eq.~(\ref{SE}) \cite{MSimon}.  
Writing
${\bf \hat{E}}$ 
in terms of the complex unit vector 
${\bf M}=({\bf N}~+~i{\bf B})/\sqrt 2$, we have 
\begin{equation}
{\bf \hat{E}}= \alpha(s) {\bf M}+\beta(s) {\bf M^*},
\end{equation}
with $|\alpha|^{2}+|\beta|^{2}=1$.

Substituting this in the Eq. ~(\ref{SE}),
it can be seen that $\alpha(s)=e^{i \chi}\alpha(0)$ and $\beta(s)=
e^{-i \chi} \beta(0)$,
where 
\begin{equation}
\chi=\int_{0}^{S}~\tau~ds,  
\label{chi}
\end{equation}
with $\tau$ as defined in Eq. (\ref{kt}). 
For a cyclic evolution, $\chi$ can be seen to be equal to the phase increase ( decrease)
of the right (left) circularly polarized light.\cite{MSimon}
Thus the phase change determined solely by the geometry of the cyclic path $\Gamma$
is fully characterized by its curvature and the torsion.
In view of the transverse nature of the electromagnetic waves, ${\bf T} \cdot {\bf \hat{E}}=0$.
This leads to a very desirable result that the dynamical phase is zero and the net phase change is equal to
$\chi$, the geometric phase.

One of the characteristics of the boundary curves of the $n$-twisted strips 
is the existence of $n$
inflexion points.
At the inflexion point, the curvature vanishes and the tangent to the curve shows a cusp.
For the M\"{o}bius strip (i.e., $n=1$), the critical boundary curve has just 
one inflexion point 
at $t=2\pi$ and $\alpha_c=4/5$. For $n=3$, the boundary curve is a trefoil,
which is knotted, and there are three inflexion points.
In general for the boundary curve of an $n$-twisted strip,
$n$ inflexion points occur at  a critical width 
$\alpha_c=1/(1+n^2/4)$,  when $\cos(nt_c/2)=-1$.\cite{PRE}
As seen in the figure ~1,
the boundary curves encode the width of the strip and the critical curve, 
namely the curve with an
inflexion point is distinguished by a vanishing of curvature around that point. 

Symbolic manipulation facilitates the determination of analytic expressions 
for the curvature $\kappa(t)$ 
and torsion $\tau(t)$, by using Eq. (\ref{ms}) in Eq. (\ref{kt}).
For the M\"{o}bius case, we get
\begin{eqnarray*}
\tau=\frac{6 \alpha [4 \cos~ t/2 + \alpha (5+8 \cos~ t) + \alpha^2 
(9 \cos~ t/2+\cos~ 3t/2)]}{\kappa^2}, 
\end{eqnarray*}
where the curvature $\kappa$ is given by\\
$\kappa = 64 + 288 \alpha \cos~ t/2 + \alpha^2 (284 + 216 \cos~ t)+
\alpha^3 (336 \cos~ t/2 + 64 \cos~ 3t/2) + \alpha^4 (65 + 54 \cos~ t+6 \cos~ 2t)$

Taylor expansion of $\kappa$ and $\tau$ near $\alpha_c$ and $t_c$ gives,
\begin{eqnarray*}
\kappa& \approx &\frac{1}{2v_c^{3/2}}[(\alpha-\alpha_c)^2+9b^2v_c^2(t-t_c)^2]^{\frac{1}{2}}\\
\tau& \approx &\frac{-3b(\alpha-\alpha_c)}{[(\alpha-\alpha_c)^2+9b^2v_c^2(t-t_c)^2]},
\end{eqnarray*}
where $b=\frac{6}{5}$ and $v_c = |d{\bf r}/dt|_c = 1/\sqrt(5)$. 
These equations show that $\tau$ 
 changes sign as the parameter $\alpha$
passes through its critical value.\cite{note}

Geometrically, the vanishing of $\kappa$ results in the divergence 
in the torsion $\tau$ at the inflexion point. However, as shown below, this singularity
is integrable and the resulting integrated $\tau$, the geometric phase is well defined.
To show this, we calculate $\chi(t)$
near the inflexion point. For arbitrary $t_0$,
the total twist in the time interval
$(-t_0+t_c,t_0+t_c)$ centered around the critical point $t_c$ is,
$\int_{t_c-t_0}^{t_c+t_0} \tau v dt = -\int_{t_c-t_0}^{t_c+t_0} \frac{b v_c (\alpha-\alpha_c)dt}{
[(\alpha-\alpha_c)^2+v_c^2b^2 (t-t_c)^2]}$.
This integral is equal to  $2 Arc \tan [\frac{t_0}{(bv_c(\alpha-\alpha_c))}]$.
As we pass through $\alpha_c$, it will change from $-\pi$ to $\pi$ ( irrespective of the value of $t_0$)
giving rise to a jump of $2\pi$. 

Figure ~3 illustrates the evolution of the geometric phase
as the light propagates through the boundary curves. 
For $\alpha < \alpha_c$, the phase factor $\chi(t)$
exhibits a wiggle near the critical point 
which becomes saw-tooth shaped at criticality. 
However, for $\alpha > \alpha_c$, the the phase change
near the critical points is smooth. As discussed above, and seen in the figure,
the polarization of light undergoes
a full rotation of $2\pi$ after passing through the critical point.
 
\begin{figure}[htbp]
\includegraphics[width=2.9in]{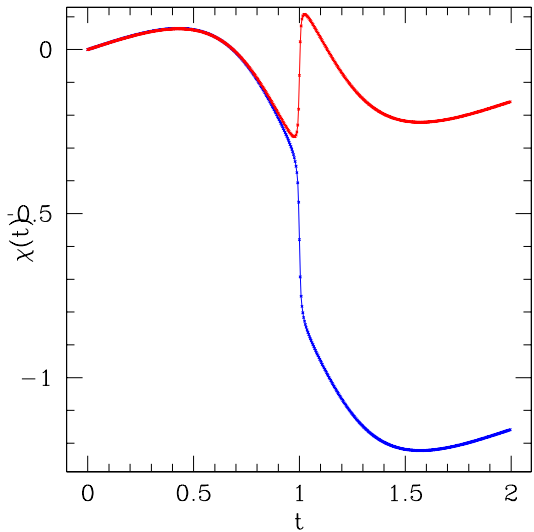}
\leavevmode
\caption{(color on line) Geometric phase (in units of $2\pi$) 
 acquired by polarized light as it passes through the closed boundary curve
 of a M\"{o}bius strip. The top and the bottom curves
 correspond respectively to strip widths just below ($\alpha=0.79$) and just above ($\alpha=0.81$)
the  critical width $\alpha_c=0.8$.}
\label{fig3}
\end{figure}

Taylor expansion of the unit normal near the inflexion point shows that
as we pass it at $t=t_c$, ${\bf N}$ rotates by $\pi$ about a fixed direction.
And for $\alpha < \alpha_c$, it rotates by $-\pi$.
The same is true also for ${\bf B}$. Therefore, the number of rotations of the
pair ${\bf N}, {\bf B}$ increases by $2\pi$ as we pass the inflection point.
For an $n$-twisted strip, the number of rotations is equal to $n$, for odd $n$.
For even $n$, there are two boundary curves. The number of rotations 
for {\it each} of them is $(n/2)$.  Thus there exists a  universal functional 
form for local geometrical quantities 
$\tau$ and $\kappa$ and a jump of $2\pi$ in the global variables 
near the inflexion point,
for the boundary curves of both orientable as well as non-orientable strips.

\begin{figure}[htbp]
\includegraphics[width=2.9in]{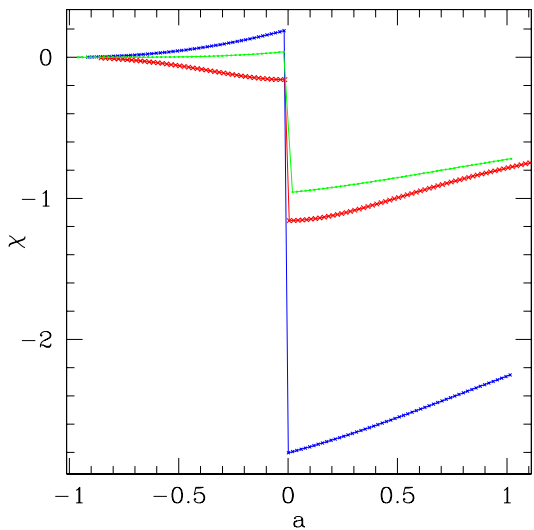}
\leavevmode
\caption{(color on line) From Top to bottom :  Geometric phase (in units of $2\pi$)
for the cyclic paths along the boundary curves of a triply-twisted strip ($n=3$), 
a doubly twisted strip $n=2$ and a 
singly-twisted, i.e., M\"{o}bius strip, for various strip widths. 
Here $a=(\alpha-\alpha_c) / \alpha_c$}
\label{fig4}
\end{figure}

\begin{figure}[htbp]
\includegraphics[width=2.9in]{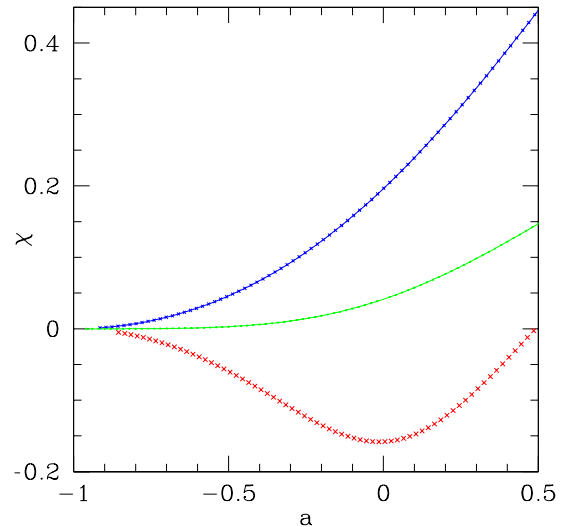}
\leavevmode
\caption{ (color on line) Same as Fig. ~ 4, except that the geometric phase is 
plotted modulo $2\pi$}.
\label{fig5}
\end{figure}

Figure ~4 shows the net phase accumulated  by light after 
propagating in a cyclic loop along the boundary curve of an  $n$-twisted 
strip, as its width varies, for $n=1,2$ and $3$. 
In figure ~5, we plot the phase modulo $2\pi$.
A striking aspect is the
characteristic minimum that we find in the phase shift in the case 
of the M\"{o}bius circuit characterized 
by the critical width of the strip. In other words, geometric phase
shows that the M\"{o}bius geometry with $n=1$ is indeed special. The absence of 
such a minimum in the boundary curves of all other (integer) $n$  values
makes the origin of such a minimum rather mysterious.

It is rather intriguing that the width for the existence of
the diabolic crossing of the eigenspectrum
of the quantum Hamiltonian coincides with the critical width resulting 
in the minimum phase shift.
It would have been natural to speculate that the global phase shift is influenced by a local
diabolo geometry. However, such a minimum exists only in the M\"{o}bius loop,
whereas the diabolic crossings characterize the boundary curves of all $n$-twisted 
strips.
Therefore, the fact that two different and somewhat unique
features occur at the same parameter is either a coincidence or it is 
conceivable that this is happening because the M\"{o}bius geometry
and topology is unique in the sense that $n=1$ is the only case where
the single boundary curve of the twisted strip is un-knotted and has a
single inflexion point. 
The presence 
of multiple inflexion points  in the $n$-twisted ribbons (with $n>1$) 
seem to play a role in such a way as to wipe out this minimum.

We conclude this paper by suggesting  possible experiments that could verify
 our theoretical results. We begin by noting that in experiments that demonstrate 
 the geometrical phase phenomenon in optics, the usual nonplanar path that has been 
 considered is either an open helix \cite{tomita}, or a helix that can be 
 effectively closed by making 
 the fiber lie along planar 
 paths (with $\tau =0$) at the two ends of the helix\cite{optics}. 
Clearly, such  a circuit has no inflexion points where $\tau$ becomes locally singular,
since $\kappa$ is a constant for a helix. To our knowledge, no experiments have 
been performed on nonplanar fibers with inflexion points. We hope that our 
theoretical results that show that the presence of inflexion points has
a nontrivial effect on the optical phase shift would motivate experimentalists
to observe it.

For example,  optical fibers which have the shape of
the boundary curve of various n-twisted strips can be perhaps be fabricated 
as follows. By twisting a sheet made of some pliable material 
such as plastic
(or any other material appropriate for optical
experiments) of some width $\alpha$ once, and gluing together the two short
edges, a closed M\"{o}bius strip can be constructed. Instead of 
winding the optical fiber on a cylinder as was
done in the experiments of Tomita-Chiao\cite{tomita} and Frins-Dultz \cite{optics}, 
 the  fiber could now  be
attached along the boundary curve  of the  above M\"{o}bius strip, and a similar
experimental setup could be used to measure the geometrical phase. The experiment 
can then be repeated
with various different widths of the M\"{o}bius, to find the dependence of the
geometrical phase on the width, with a characteristic minimum phase 
occurring at a critical
width. Similar measurements can be done with optical fibers in
the shape of the boundary curve of a strip twisted $n$ times, and repeated
for different widths. 

Finally, our studies may be relevant in optical fibers used 
in tuning of polarization, as such a phase can be seen 
in low birefringent optical fibers\cite{tune}.
We hope that our studies will also stimulate laboratory research 
involving lasers and condensates in this novel class of twisted 
geometries that we have studied.\\

RB thanks the Council of Scientific and Industrial Research, India, for
financial support under the Emeritus Scientist Scheme.


\begin{references}
\bibitem{note} We would like to point out that
the functional form of $\kappa$ and $\tau$ near the inflexion point is identical to that obtained 
in an earlier study\cite{Ricca}
investigating the generic behavior of a curve in the vicinity of an 
inflexion point, the only difference being that the value of $b$ was unity
there.
\bibitem{fn} However, note that this circuit cannot be regarded as the boundary 
curve of an $n$-twisted strip any more, since $n$ is fractional.
\end{references}
\end{document}